\documentclass[twocolumn,nofootinbib,showpacs,prd,superscriptaddress]{revtex4}

\usepackage{graphics,graphicx}

\usepackage{amsmath}
\usepackage{psfrag}
\usepackage{dcolumn}
\usepackage{psfrag}

\newcommand{\beq}{\begin{equation}}
\newcommand{\eeq}{\end{equation}}
\newcommand{\bea}{\begin{eqnarray}}
\newcommand{\eea}{\end{eqnarray}}

\newcommand{\vek}[1]{\boldsymbol{#1}}

\def\no{\nonumber \\}

\allowdisplaybreaks

\begin{document}

\title{	
Comparison between numerical-relativity and post-Newtonian waveforms from spinning binaries: 
the orbital hang-up case
}

\author{Mark Hannam} \affiliation{Theoretical Physics Institute, University of
  Jena, 07743 Jena, Germany} 
\author{Sascha Husa} \affiliation{Max-Planck-Institut f\"ur
  Gravitationsphysik, Albert-Einstein-Institut, Am M\"uhlenberg 1, 14476 Golm,
  Germany}  
\author{Bernd Br\"ugmann} \affiliation{Theoretical Physics Institute,
  University of Jena, 07743 Jena, Germany}
\author{Achamveedu Gopakumar} \affiliation{Theoretical Physics Institute, University of
  Jena, 07743 Jena, Germany} 

\date{\today}

 \begin{abstract}
We compare results from numerical simulations of spinning binaries in the
``orbital hangup'' case, where the binary completes at least nine orbits
before merger, with post-Newtonian results using the approximants TaylorT1,
T4 and Et. We find that, over the ten cycles before the gravitational-wave
frequency reaches $M\omega = 0.1$, the accumulated phase disagreement between
NR and 2.5PN results is less than three radians, and is less than 2.5 radians
when using 3.5PN results. The amplitude disagreement between NR and
restricted PN results increases with the black holes' spin, from about 6\% in
the equal-mass case to 12\% when the black holes' spins are $S_i/M_i^2 = 0.85$.
Finally, our results suggest that the merger waveform will play an important
role in estimating the spin from such inspiral waveforms.
\end{abstract}

\pacs{
04.20.Ex,   
04.25.Dm, 
04.30.Db, 
95.30.Sf    
}

\maketitle

\section{Introduction}

Direct observation of gravitational waves may be at hand.  Several
ground-based gravitational-wave (GW) detectors have reached design sensitivity
~\cite{Waldman06,GEOStatus:2006,Acernese2006}, and the recent completion of the
LIGO S5 science run, where science data have been taken for almost two years,
marks an important milestone in the field. With the joining of the GEO600
and Virgo detectors for part of this run, an important step has also been taken
in establishing a global network of detectors, and data from all these detectors are
currently being analyzed. 

For black-hole binary systems,
which are the strongest expected GW sources, current GW template banks
use waveforms calculated by analytic post-Newtonian (PN) approximation methods
to model waves from a binary's slow inspiral, and perturbation-theory
techniques to model the ringdown of the final merged black hole. It is not clear
how well these models  
describe the crucial merger phase, which will produce the strongest signal. 
Recent breakthroughs in numerical relativity (NR) \cite{Pretorius:2005gq,
Campanelli:2005dd,Baker05a} have made it possible to describe the late
inspiral, merger and ringdown in full general relativity, and future GW searches
will be able to use ``hybrid'' waveforms produced by combining PN and NR
waveforms \cite{Pan:2007nw,Ajith:2007qp,Ajith:2007kx,Buonanno:2007pf}.

The construction of hybrid waveforms requires an understanding of the frequency
range in which PN and NR waveforms overlap, and how close to merger the
PN waveforms cease to be sufficiently accurate. Studies so far have focused on
nonspinning binaries 
\cite{Buonanno:2006ui,Baker:2006ha,Pan:2007nw,Hannam:2007ik,Boyle:2007ft}. 
The broad conclusion from that work was that up to a few orbits before
merger, standard PN approximants predict the phase evolution of the binary to within an 
accuracy of about 1 radian \cite{Baker:2006ha,Hannam:2007ik,Boyle:2007ft},
the lowest-order PN contribution to the wave amplitude
disagrees with numerical results by about 6\% over the frequency range considered
in numerical simulations, and the error from higher-order PN amplitude corrections 
is reduced to only a few percent up to
10-15 orbits before merger \cite{Hannam:2007ik,Boyle:2007ft}. 

We now consider binaries in which the individual black holes are
spinning. Spinning binaries have already 
been simulated in many contexts 
\cite{Campanelli:2006uy,Campanelli:2006fy,Campanelli:2007ew,Campanelli:2007cg,Lousto:2007db,Koppitz-etal-2007aa,Pollney:2007ss,Rezzolla:2007xa,Rezzolla:2007rd,Herrmann:2007ac,Herrmann:2007ex,Baker:2007gi,Gonzalez:2007hi,Brugmann:2007zj,Berti:2007nw,Marronetti:2007wz,Tichy:2007hk},
but in this work we present the first {\it long} ($>15$ GW cycles) simulations of
spinning binaries, and compare with PN results during the inspiral. 
 As a first example from the large
parameter space, we consider equal-mass binaries with equal spins that
are parallel to the orbital angular momentum of the binary. In addition, we compare
the spinning waveforms with their nonspinning counterparts, and begin to 
address the question of how well the spin can be estimated from an observation
of one of these waveforms. We find that, during the slow inspiral stage,
spinning and nonspinning binaries are difficult to distinguish; 
while during the merger and ringdown the waveforms can be clearly distinguished.

We summarize our numerical methods in Section~\ref{sec:numerics}, and also the
main features of the numerical simulations that we performed, and in
Section~\ref{sec:PNmethods} summarize three PN calculations of a spinning
binary's phase evolution, the Taylor T1, T4 and Et methods. In
Section~\ref{sec:pncomparison} we compare the NR phase evolution and amplitude
from the $(l=2,m=2)$ mode of $r\Psi_4$ with the PN phase calculated from the
TaylorT1, T4 and Et approximants, and with the PN amplitude at restricted
(quadrupole) order. We also compare the numerical waveforms with each other
in Section~\ref{sec:numcomparison} to get an estimate of how much nonspinning
and spinning waveforms differ in the inspiral and merger phases.

\section{Numerical methods and results}
\label{sec:numerics}

We performed numerical simulations with the BAM code
\cite{Bruegmann:2006at,Bruegmann2004,Husa2007a}. 
The code starts with black-hole binary puncture initial data 
\cite{Brandt97b,Bowen80} generated using a pseudo-spectral code
\cite{Ansorg:2004ds}, and evolves them with the $\chi$-variant of the
moving-puncture \cite{Campanelli2006,Baker2006} version of the BSSN
\cite{Shibata95,Baumgarte99} formulation of the 3+1 Einstein 
evolution equations. Spatial finite-difference derivatives are
sixth-order accurate in the bulk \cite{Husa2007a}, Kreiss-Oliger
dissipation terms converge at fifth order, and a fourth-order Runge-Kutta
algorithm is used for time evolution. 
The gravitational waves emitted by the binary are calculated from the
Newman-Penrose scalar $\Psi_4$, and the details of our implementation of
this procedure are given in \cite{Bruegmann:2006at}.

The new simulations we performed for this analysis are summarized in 
Tables~\ref{tab:simulations} and \ref{tab:parameters}. All simulations are of
equal-mass binaries; the black-hole punctures are placed on the $y$-axis
at $y = \pm D/2$, and given momenta in the $x$ direction of $p_x = \mp p$. 
The black holes each have the same spin angular momentum $S_i$ oriented 
parallel to the total orbital angular momentum, i.e., in the positive $z$ direction. 
The grid setup is described following the notation introduced in \cite{Bruegmann:2006at}. 
For example, $\chi_{\eta=2}[5\times 64:5\times 128:6] $  indicates that the 
simulation used the $\chi$ variant of the moving-puncture method, five nested
mesh-refinement boxes with a base value of $64^3$ points surround each black hole,
and five nested boxes with $128^3$ points surround the entire system, and there are
six mesh-refinement buffer points. 
As summarized in Table~\ref{tab:parameters}, simulations were performed with 
spins $S/M_i^2 = 0.25,0.5,0.75,0.85$. We also make use of the results from
nonspinning binaries, i.e., $S_i/M_i^2 = 0$, as reported in \cite{Hannam:2007ik}.

\begin{table}
\caption{\label{tab:simulations}
Summary of grid setup for numerical simulations. The grid parameters follow the
notation introduced in \cite{Bruegmann:2006at}; see text. $h_{min}$ denotes the resolution on the 
finest level and $h_{max}$ the resolution on the coarsest level. The outer 
boundary of the computational domain is at approximately $r_{max}$. 
The simulations with spin $S/M_i^2 = 0.75,0.85$ use one extra level of mesh 
refinement. 
} 
\begin{ruledtabular}
\begin{tabular}{l|r|r|r}
Run & $h_{min}$  & $h_{max}$ & $r_{max}$  \\[10pt]
\hline
\multicolumn{4}{l}{$S/M_i^2 = 0,0.25,0.5$ simulations }\\
\hline
$\chi_{\eta=2}[5\times 64:5\times 128:6] $ & $M/42.7$ &  $12 M$ & $774M$ \\
$\chi_{\eta=2}[5\times 72:5\times 144:6] $ & $M/48.0$ &  $32/3 M$ & $773M$ \\
$\chi_{\eta=2}[5\times 80:5\times 160:6] $ & $M/53.3$ &  $48/5 M$  & $773M$ \\
\hline
\multicolumn{4}{l}{$S/M_i^2 = 0.75, 0.85$ simulations }\\
\hline
$\chi_{\eta=2}[6\times 64:5\times 128:6] $ & $M/85.3$ &  $12 M$ & $774M$ \\
$\chi_{\eta=2}[6\times 72:5\times 144:6] $ & $M/96.0$ &  $32/3 M$ & $773M$ \\
$\chi_{\eta=2}[6\times 80:5\times 160:6] $ & $M/106.7$ &  $48/5 M$  & $773M$ \\
\end{tabular}
\end{ruledtabular}
\end{table}

The physical parameters are given in Table~\ref{tab:parameters}. As the spin
is increased, the mass parameter $m_i$ decreases. This is partly because this
quantity parametrizes the mass associated with the area of the apparent
horizon, $M_{AH} = \sqrt{A / 16 \pi}$, where $A$ is the area of the horizon,
while the total mass of the black hole is estimated by a variant of the
Christodoulou formula  \cite{Christodoulou70}, \begin{equation} 
M_i^2 = (M_{AH,i})^2 + \frac{S_i^2}{4 (M_{AH,i})^2}. \label{BHmass}
\end{equation} If we are to keep $M_i$ constant as $S_i$ is increased, then
$M_{AH}$ must decrease, and therefore $m_i$ decreases. The decreasing value of
$m_i$ is also due to the extra ``junk'' energy that the spin adds to the
initial data: the Bowen-York extrinsic curvature used in the initial-data
construction contains unwanted gravitational radiation that increases in
amplitude as the spin is increased. That radiation adds to the mass of the
black hole, and can be compensated by further lowering $m_i$, but only up to a
point: eventually there is too much junk radiation, and we reach a limit in
the $S_i/M_i^2$ that we can obtain. This limit has been found experimentally
to be about $S_i/M_i^2 \approx 0.928$ \cite{York-Piran-1982-in-Schild-lectures,Choptuik86b}. 
A form of puncture data that permits
higher spins has been suggested \cite{Hannam:2006zt}, but for the present
study we consider spins no larger than $S_i/M_i^2 = 0.85$, which is well below
the limit for Bowen-York data. 

 The initial momenta for quasi-circular inspiral were calculated using a
 2.5-PN-accurate procedure based on the results in \cite{Kidder1995} and
 outlined in \cite{Brugmann:2007zj}. We expect these parameters to lead to
 inspiral with a small eccentricity, and this is indeed what we see in the
 numerical data; the eccentricity is typically of the order of $e \approx
 0.006$.  We found that the procedure described  in \cite{Husa:2007ec} to
 produce lower-eccentricity inspiral, although applicable for spinning
 binaries, does not yet include sufficient accuracy in the spin terms to be of
 use for the scenarios described in this work.

\begin{table}
\caption{\label{tab:parameters}
Physical parameters for the moving-puncture simulations: the coordinate
separation, $D/M$, the mass parameters in the puncture data construction,
$m_i/M$, and the momenta $p_x/M$. The punctures are placed on the $y$-axis, 
and for all
simulations the total initial black-hole mass is $M = 1$.} 
\begin{ruledtabular}
\begin{tabular}{l|r|r|r|r}
Simulation & $S_i/M_i^2$ & $D/M$ & $m_i/M$  & $p_x/M$     \\
\hline 
S25   & 0.25           & 12.0  & 0.47579  & 0.083813 \\
S50   & 0.50           & 11.0  & 0.43277  & 0.087415 \\
S75   & 0.75           & 10.0  & 0.33608  & 0.091435 \\
S85   & 0.85           & 10.0  & 0.25628  & 0.090857 \\
\end{tabular}
\end{ruledtabular}
\end{table}

We Richardson extrapolate our data with respect to numerical resolution and radiation
extraction radius as described in \cite{Hannam:2007ik}. We first split the $(l=2,m=2)$ mode
of the waveform
$r \Psi_4$ into amplitude and phase according to \begin{equation}
r \Psi_4 = A(\phi(t)) e^{-i\phi(t)}, \label{eqn:AmpPhase}
\end{equation} The amplitude is in turn written as a function of phase, $A(\phi)$, and
this function is Richardson extrapolated with respect to numerical resolution. Five such
functions are produced, one for each extraction radius, $R_{ex} = \{ 50,60,70,80,90\}M$. 
We then note that the dependence of the amplitude on extraction radius is modeled well
by \begin{equation}
A(\phi,R_{ex}) = A_{\infty}(\phi) + \frac{k(\phi)}{R_{ex}^2} + O\left(
  \frac{1}{R_{ex}^3} \right),
\label{eqn:AmpFalloff} 
\end{equation} and applying a curve fit of the form (\ref{eqn:AmpFalloff}) to $A(\phi)$, 
we estimate $A(\phi_,R_{ex} \rightarrow \infty)$. Although a method has been
suggested to also extrapolate the phase to infinite extraction radius
\cite{Boyle:2007ft}, in this work we again follow the procedure described in
\cite{Hannam:2007ik} and use the phase from the largest extraction radius,
$R_{ex} = 90M$.

In the equal-mass case studied in \cite{Hannam:2007ik}, we found that the
numerical results were cleanly sixth-order convergent, and were therefore able
to both remove the sixth-order error term by Richardson extrapolation, and to
estimate the uncertainty due to higher-order error terms. The present
simulations of spinning binaries do not show such clean convergence. All of
the simulations exhibit convergence between fifth and sixth order. Since the
code contains fourth, fifth and sixth-order elements, it is not obvious at
which resolutions each error term will dominate, and we make the most
conservative choice of using the highest-resolution results and estimating an
uncertainty by assuming only fourth-order convergence. Even with this
conservative estimate of the discretization error, we find that, as in the
equal-mass case, the errors are anyway dominated by the finite extraction
radii. In general we estimate the uncertainty in our waveform amplitude as
less than 3\%, and in the phase our uncertainty is 0.25 radians over the
frequency range that we will consider for comparison with PN results. 

These uncertainty estimates do not take into account the effect of eccentricity.
In the nonspinning case studied in \cite{Hannam:2007ik,Gopakumar:2007vh}, the
eccentricity was $e < 0.0016$, and the error in the phase evolution due to 
the eccentricity was estimated as being well below the finite-difference and
finite-extractdion-radii errors. Figure~15 of \cite{Hannam:2007ik} shows
that the accumulated phase error of a simulation with $e \sim 0.008$ is around
0.2 radians, which {\it is} comparable to the numerical phase error. In the 
simulations presented in this work, the eccentricity can be as high at $e \sim 0.006$,
and so significant eccentricity-induced phase errors may be expected. 
Based on Figure~15 in \cite{Hannam:2007ik}, we estimate that such errors 
are no larger than 0.2 radians. This is a systematic error: the binary merges
sooner if the eccentricity is increased, and therefore the accumulated phase
disagreement $\Delta \phi_e = \phi_{e=0} - \phi_{e>0}$ will always be positive.
This systematic uncertainty should be taken into account in the comparisons we perform
with PN approximants in Section~\ref{sec:pncomparison}. The effect of the 
eccentricity on the wave amplitude is to produce oscillations in the amplitude. 
For the simulations with higher spin, the eccentricity is larger, and the
resulting oscillations are larger. This is clear in Figure~\ref{fig:AmplitudeComparison},
where the oscillations are visible in the highest-spin case shown ($S_i/M_i^2 = 0.75$),
and the highest-spin results are not shown because the oscillations reduce
the clarity of the figure. We emphasize that the errors due to eccentricity
may appear from our results to be larger for systems with larger spin, but this 
is not necessarily the case: the errors are simply larger when the eccentricity
is large, and in the particular simulations we have done the larger-spin 
configurations also have larger eccentricity, and therefore larger
eccentricity-induced errors.

Some of the features of the simulations are summarized in
Table~\ref{tab:numresults}: the mass and spin of the final black hole,
and two quantities that allow comparison between
simulations, and which demonstrate 
the orbital hangup effect first observed in numerical simulations in
\cite{Campanelli:2006uy}. Starting from a GW frequency of $M\omega = 0.06$, we
list the time it takes the GWs to reach their maximum amplitude, $\Delta t_A$, and the
number of GW cycles up to the maximum. (Dividing this number by
two gives the corresponding number of the binary's orbits.) The numbers
clearly show that the binary's merger is ``hung up'' by the presence of spin. 

The mass of the final black hole was estimated by subtracting the radiated
energy, as measured at the largest extraction radius $R_{ex} = 90M$, from the
ADM mass of the spacetime, as calculated on the initial timeslice. The spin of
the final black hole was estimated by comparing a fit of the ringdown with
analytic quasinormal mode results. The values for the mass and spin are
consistent with those already given in the literature
\cite{Campanelli:2006uy,Marronetti:2007wz,Rezzolla:2007xa}.

\begin{table}
\caption{\label{tab:numresults}
Selected global features of the simulations: the mass and spin of the final
black hole, $M_{\rm final}$ and $S_{\rm final}/M_{\rm final}^2$; the time $\Delta t_A$ 
(in units of $M$) for the GW to evolve from $M\omega = 0.06$ to its maximum amplitude; 
and the number of GW
cycles $\Delta N_{GW}$ between $M\omega = 0.06$ and the amplitude maximum. 
} 
\begin{tabular}{||l|r|r|r|r|}
\hline
Simulation & $M_{\rm final}$ & $S_{\rm final}/M_{\rm final}^2$ & $\Delta
t_A (M)$ & $\Delta N_{GW}$     \\
\hline 
D12   & 0.950 & 0.680  &  719 & 11.0 \\
S25   & 0.943 & 0.757 &  819 & 12.7 \\
S50   & 0.932 & 0.826 &  917 & 14.7 \\
S75   & 0.920 & 0.896 &  1040 & 16.8 \\
S85   & 0.911 & 0.918 &  1096 & 18.0 \\
\hline
\end{tabular}
\end{table}

\begin{figure}[t]
\centering
\includegraphics[height=4cm]{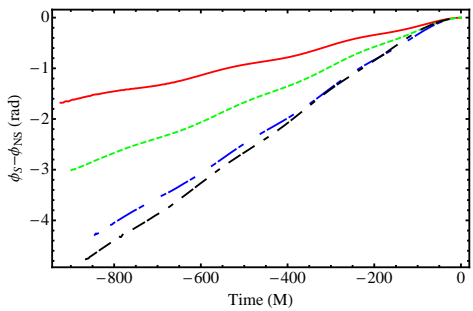}
\caption{The accumulated phase difference between nonspinning and spinning
  binaries for the ten cycles before $M\omega = 0.1$, aligned such that the 
  phase disagreement is zero when $M\omega = 0.1$. The black-hole spins
  are, in order of increasing magnitude of the phase disagreement, $S_i/M_i^2 =
  0.25,0.5,0.75,0.85$. Note that the
  differences are in some cases comparable to the phase disagreements with PN results 
  that we will present in Section~\ref{sec:pncomparison}.} 
\label{fig:NumPhaseDifference}
\end{figure}

Figure~\ref{fig:NumPhaseDifference} shows the accumulated phase error between
the nonspinning binary simulations presented in \cite{Hannam:2007ik}, and the
various spin cases, for the ten cycles before the given spin waveform reaches a GW
frequency of $M\omega = 0.1$. The GW phases are aligned such that the phase
difference is zero when $M\omega = 0.1$ and we also relabel the time so that $t=0$
at this point; this procedure is described in more detail
in Section~\ref{sec:phase}. We see that for $S_i/M_i^2 = 0.25$, the accumulated
phase difference  is about 1.6
radians, while for $S_i/M_i^2 = 0.85$, the accumulated phase difference is
almost five radians. These differences should be borne in mind when we compare
with the post-Newtonian phase predictions.

\section{Various prescriptions for GW phase evolutions in PN relativity}
\label{sec:PNmethods}

 In this section, we provide formulae, extracted from
 Refs.~\cite{Damour:2001bu,Blanchet:2001ax,Blanchet:2004ek,Kidder1995,Poisson:1997ha,Alvi:2001mx,Blanchet:2006gy,Faye:2006gx}, required to  construct TaylorT1, T4 and Et templates for  
inspiraling equal-mass binary black holes, having their spins ${\vek S_1} $
and $\vek{S_2}$ aligned with the Newtonian angular momentum unit vector $\vek{l}$.
(We retain $G$ and $c$ in the formulas in this section, although for the
remainder of the paper we will adopt geometrized units $G = c = 1$.)
The aim is to provide GW phase evolutions under TaylorT1, T4 and Et prescriptions
that include all the non-spinning contributions to 3.5PN order and 
the spin effects to 2.5PN order. We require, as in the case of non-spinning compact binaries,
the dynamical (orbital) energy ${\cal E}(x)$,
expressed as a PN series in terms
of $x = \left ( G\, M\omega_b /c^3 \right )^{2/3}$,
$ \omega_b(t) $ being the binary's orbital angular frequency,
and the GW energy luminosity
$ {\cal L }(x)$. 
The fact that we are dealing with compact binaries having aligned spins implies that
to the highest PN order considered in this paper,
the spin vectors $\vek S_1$ and $\vek S_2$ have their
directions and magnitudes fixed.

   The 3.5PN accurate $ {\cal L }(x)$ and 3PN accurate ${\cal E}(x)$ associated
with non-spinning comparable mass compact binaries are extractable from
Refs.~\cite{Damour:2001bu,Blanchet:2001ax,Blanchet:2004ek}. 
The lowest-order General Relativistic spin-orbit coupling, appearing at 1.5PN order for 
maximal Kerr black holes, and  spin-spin effects entering at 2PN order 
make contributions to $ {\cal L }(x)$  and ${\cal E}(x)$ at relative
1.5PN and 2PN orders \cite{Kidder1995}. 
In the usual terminology adapted in the PN literature, this implies that
spin-orbit coupling 
provides corrections to ${\cal E}(x)$ in terms of $ x^{3/2} $ with respect to 
its Newtonian counterpart  and a similar rule applies for  $ {\cal L }(x)$.
The monopole-quadrupole interactions
affect ${\cal L }(x)$  and ${\cal E}(x)$ at relative 
2PN order \cite{Poisson:1997ha}.
The next to leading order General Relativistic spin-orbit coupling affects 
${\cal L }(x)$  and ${\cal E}(x)$ at relative  2.5PN order
\cite{Blanchet:2006gy,Faye:2006gx}.   
However, for inspiraling spinning black holes, 
there exist corrections to  ${\cal L }(x)$, also 
appearing at the relative 2.5PN order, 
due to the energy flowing in to the black holes. For comparable mass spinning black holes,
the above-mentioned contributions are derived in Ref.~\cite{Alvi:2001mx}.
For PN computations, available in
Refs.~\cite{Damour:2001bu,Blanchet:2001ax,Blanchet:2004ek,Kidder1995,Poisson:1997ha,Blanchet:2006gy,Faye:2006gx},  
a non-spinning black hole is modeled as a point-particle 
(spherically symmetric
mass distribution), and
a spinning black hole is approximately
treated as a spinning spherically symmetric mass
distribution.
It is important to note that properties of black holes, like
presence of black hole horizons, play no role in
these PN computations.
Due to inclusion of black hole absorption effects, available in Ref.~\cite{Alvi:2001mx},
it is reasonable to argue that our construction of fully 2.5PN accurate 
TaylorT1, T4 and Et templates
are really applicable for spinning black hole binaries.

We define spins 
as $\vek {S_i} = G\, m_i^2\, \chi_i\, \vek{s_i}, i =1,2 $ and in this paper, we 
impose the restrictions, $ \vek {s_i} \cdot \vek {l} = +1 $ 
and $ \vek s_1 \cdot \vek s_2 = +1 $.
For compact binaries having aligned spin configurations, it is also possible to
employ the restricted PN waveforms
\begin{align}
h(t) \propto  x(t)^{2/3} \, \cos 2\, \phi(t)\,,
\label{EqS1}
\end{align}
To obtain the GW phase evolution $\phi(t)$ in the TaylorT1 approximant,
one  numerically solves the following two differential equations:
\begin{subequations}
\label{EqS2}
\begin{align}
\label{EqS2a}
\frac{d \phi (t)}{dt} \equiv \omega_b (t) &= \frac{c^3}{G\,M}\, x^{3/2}\,,\\
\frac{d\,x(t)}{dt} &=  -{\cal L}(x) \left( \frac{ d {\cal E}}{d x}\right)^{-1}\,.
\label{EqS2b}
\end{align}
\end{subequations}
 
 Therefore, to obtain the GW phase evolution, relevant for equal mass 
black hole binaries with aligned spins, 
under the TaylorT1 approximant
that includes all the 3.5PN accurate non-spinning and 2.5PN accurate 
spin effects, we require the
following expressions for  ${\cal L}(x)$ and ${\cal E}(x)$.  
\begin{subequations}
\label{EqP3}
\begin{align}
\label{EqP3a}
{\cal L}(x) &= {\cal L}(x)|_{\rm NS} + {\cal L}(x)|_{\rm S} \,,
\\
{\cal L}(x)|_{\rm NS} &=
{\frac {2\,{c}^{5}\,}{5\,G}}\, x^5\, \biggl \{
1-{\frac {373}{84}}\,x+4\,\pi\,{x}^{3/2}-{\frac {59}{567}}\,{x}^{2}
\no
&
-{
\frac {767}{42}}\,\pi\,{x}^{5/2}
+ \biggr [
 {\frac {18608019757}{
209563200}}+{\frac {355}{64}}\,{\pi }^{2}
-{\frac {1712}{105}}\,\gamma
\no
&
-{\frac {1712}{105}}\,\ln  \left( 4\,\sqrt {x} 
\right) 
 \biggr ] {x}^{3}
+{\frac {16655}{6048}}\,\pi \,{x}^{7/2}
\biggr \}\,,
\\
{\cal L}(x)|_{\rm S} &=
{\frac {2\,{c}^{5}\,}{5\,G}}\, x^5\, \biggl \{
- \left(\chi_{{1}}+\chi_{{2}} \right) {x}^{3/2}
+ \frac{1}{2} \,
\biggl ( {\chi_{{1}}}^{2}+{\chi_{{2}}}^{2}
\no
&
+{\frac {31}{16} }\,\chi_{{1}}\chi_{{2}} \biggr ) {x}^{2}
+ 
\biggl [
{\frac {799}{504}}\, \left ( \chi_{{1}} + \chi_{{2}} \right )
\no
&
- \frac{1}{32}\, \biggl(
 \chi_{{1}}+\chi_{{2}}+3\,
{\chi_{{1}}}^{3}+3\,{\chi_{{2}}}^{3} 
\biggr ) \nu
\biggr ]\,  {x}^{5/ 2}
\biggr \}\,,
\\
{\cal E}(x) &=  {\cal E}(x)|_{\rm NS} + {\cal E}(x)|_{\rm S} \,,\\
{\cal E}(x)|_{\rm NS}  &= -\frac{M\,c^2}{8}\,x \biggl \{ 
1+ 
-{\frac {37}{48}}\,x
-{\frac {1069}{384}}\,{x}^{2}
\no
&
+ \biggl [ {\frac {1427365}{331776}}
-{\frac {205}{384}}\,{\pi}^{2}
  \biggr ]  {x}^{3}
\biggr \}\,, \\
{\cal E}(x)|_{\rm S}  &= -\frac{M\,c^2}{8}\,x \biggl \{ 
\frac{7}{6}\, \left( \chi_{{1}}+ \chi_{{2}} \right) {x}^{3/2}
-\frac{1}{4}\, \left( \chi_{{1}}+ \chi_{{2}} \right)^2\, {x}^{2}
\no
&
+ {\frac {335}{144}}\, \left( \chi_{{1}}+\chi_{{2} } \right) {x}^{5/2}
\biggr \}\,,
\label{EqP3b}
\end{align}
\end{subequations}
where $\gamma$ is the Euler gamma, and where we use $\nu$ to mark
contributions due to black-hole absorption effects; it is set to one to take
those effects into account.

 The TaylorT4 approximant is obtained by Taylor expanding the right hand side 
of Eq.~(\ref{EqS2b}) for $dx/dt$
and truncating it at the appropriate reactive PN order.
Therefore, to construct GW phase evolution in the TaylorT4 approximant
that contains  
all the 3.5PN accurate non-spinning and 2.5PN accurate
spin effects, 
the following
set of differential equations are numerically integrated:
\begin{subequations}
\label{EqP4}
\begin{align}
\label{EqP4a}
\frac{d \phi (t)}{dt} &\equiv \omega_b (t) = \frac{c^3}{G\,M}\, x^{3/2}\,,\\
\frac{d\,x(t)}{dt} &=  \frac{d\,x(t)}{dt} |_{\rm NS} +  \frac{d\,x(t)}{dt} |_{\rm S}\,,\\
\frac{d\,x(t)}{dt} |_{\rm NS} &=
\frac{16\,c^3}{5\,G\,M}\, x^5\, \biggl \{
1
-{\frac {487}{168}}\,x
+4\,\pi\,{x}^{3/2}
\no
&
+{ \frac {274229}{72576}}\,{x}^{2}
-{\frac {254}{21}}\,\pi\,{x}^{5/2}
\no
&
+
\biggl [
{\frac {178384023737}{
3353011200}}-{\frac {1712}{105}}\,\gamma+{\frac {1475}{192}}\,{\pi}^{2
}
\no
&
-{\frac {856}{105}}\,\ln \left( 16\,x \right)  
\biggr ]
 {x}^{3}
+{\frac {3310}{189}}\,\pi\,{x}^{7/2}
\biggr \}
\,,\\ 
\frac{d\,x(t)}{dt} |_{\rm S} &=
\frac{16\,c^3}{5\,G\,M}\, x^5\, \biggl \{
-{\frac {47}{12}}\, \biggl [ \chi_{{1}} + \chi_{{2}} \biggr ]\, {x}^{3/2}
\no
&
+ \biggl [ 
\frac{5}{4}\,\left (  {\chi_{{1}}}^{2} + {\chi_{{2}}}^{2} \right )
+ {\frac {79}{ 32}}\,\chi_{{1}}\chi_{{2}}
\biggr ]\,{x}^{2}
\no
&
+ \biggl [
-{\frac {1}{32}}\, \left ( \chi_{{1}} + \chi_{{2}}
+ 3\, {\chi_{{1}}}^{3} + 3\,{\chi_{{2}}}^{3} \right )\, \nu
\no
&
-{\frac {8347}{2016}}\, \left ( \chi_{{1}} + \chi_{{2}} \right )
\biggr ]\,{x}^{5/2}
\biggr \} \,.
\label{EqP4b}
\end{align}
\end{subequations}

  The construction of the TaylorEt approximant  
requires PN accurate expressions for $\omega_b$ in terms of ${\cal E}$. 
The radiation reaction induced inspiral is incorporated by
expressing 
${\cal L}$ in terms of ${\cal E}$.
The restricted PN waveform associated with the
TaylorEt approximant reads
\begin{align}
\label{EqP5}
h(t) & \propto {\cal E}(t) \, \cos 2\,\phi (t) \,,
\end{align}
The PN accurate temporal evolutions for $\phi (t) $ and ${\cal  E}(t)$, 
that includes
all the 3.5PN accurate non-spinning and 2.5PN accurate
spin effects,
are obtained by solving the following coupled differential equations.
\begin{subequations}
\label{EqP6}
\begin{align}
\label{EqP6a}
\frac{d \phi (t)}{dt} &\equiv \omega_b (t) =  
 \omega_b(t)|_{\rm NS} + \omega_b(t)|_{\rm S} \,,\\
\omega_b(t)|_{\rm NS} &=
\frac{ c^3}{G\,M}\,\xi^{3/2} \biggl \{
1
+{\frac {37}{32}}\, \xi
+{\frac {12659}{2048}}\,{\xi}^{2}
\no
&
+ \biggl [
 {\frac {205}{256}}\,{\pi}^{2}+{\frac {3016715}{196608}}
\biggr ]
{\xi}^{3}
\biggr \}\,,
\\
\omega_b(t)|_{\rm S} &=
\frac{ c^3}{G\,M}\,\xi^{3/2} \biggl \{
-\frac{7}{4} \biggl [ \chi_{{1} } + \chi_{{2}} \biggr ] {\xi}^{3/2}
\no
&
+ \frac{3}{8} \biggl [ \left (\chi_{{1} } + \chi_{{2}} \right )^2 \biggr ] {\xi}^{2}
- {\frac {655}{64}} \, \biggl [ \chi_{{1}} + \chi_{{2}} \biggr ]
\,{\xi}^{5/2}
\biggr \}\,,\\
\frac{d\,\xi (t)}{dt} &= \frac{d\,\xi (t)}{dt} |_{\rm NS} + 
 \frac{d\,\xi (t)}{dt} |_{\rm S}\,,\\
\frac{d\,\xi (t)}{dt} |_{\rm NS} &=
{\frac {16\,c^3}{5\,G\,M}}\,{\xi}^{5} 
\biggl \{
1
-{\frac {197}{336}}\,\xi
+4\,\pi\,{\xi}^{3/2}
\no
&
+{\frac {374615}{72576}} \,{\xi}^{2}
+{\frac {299}{168}}\,\pi\,{\xi}^{5/2}
+ \biggl [ 
{\frac {3155}{384 }}\,{\pi}^{2}
\no
&
-{\frac {1712}{105}}\,\ln  \left( 4\,\sqrt {\xi} \right) 
+{\frac {4324127729}{82790400}}
\no
&
-{\frac {1712}{105}}\,\gamma 
\biggr ] { \xi}^{3}
+{\frac {4155131}{96768}}\,\pi\,{\xi}^{7/2}
\biggr \}
\,,\\
\frac{d\,\xi (t)}{dt} |_{\rm S} &=
{\frac {16\,c^3}{5\,G\,M}}\,{\xi}^{5} 
\biggl \{ 
- {\frac {41}{6} } \biggl [  \chi_{{1}}+\chi_{{2}} 
\biggr ] {\xi}^{3/2}
\no
&
+ 
\biggl [ \frac{7}{4} \, \left ( \chi_1^2 + \chi_2^2 \right )
+{\frac {111 }{32}}\,\chi_{{1}}\chi_{{2}}  
\biggr ] {\xi}^{2}
\no
&
+ 
\biggl [ -\frac{1}{32}
\left( 
\chi_{{1}}+\chi_{{2}} + 3\, \chi_1^{3} + 3\, \chi_2^{3}
\right) \nu
\no
&
-{\frac {9943}{448}}\,\left ( \chi_{{1}} + \chi_{{2}} \right )
\biggr ]\, {\xi} ^{5/2}
\biggr \}\,,
\label{EqP6b}
\end{align}
\end{subequations}
where 
$\xi = -{2\, \cal E}/\mu\,c^2$ and $\mu$ is the reduced mass.
In this paper 
we keep $d \phi/dt$ 
to its highest PN order and  change PN orders of 
$d \xi/dt $ to create various PN accurate TaylorEt approximants.
The values of $\xi $ corresponding to  any initial and final GW 
frequencies can be numerically evaluated using the right-hand side of Eq.~(\ref{EqP6a}).
This is possible due to the fact that for GWs from spinning compact binaries, having
negligible orbital eccentricities,
the frequency of the dominant harmonic is $ f_{\rm GW} \equiv \omega_b/2 $.

 We want to emphasize that the above mentioned Taylor approximants can only provide,
from a strict PN point of view, 
fully 2.5PN accurate inspiral phase evolution for spinning black-hole 
binaries. This is because there are yet to be 
computed spin effects appearing at conservative and reactive 
3PN and 3.5PN orders. 
Another point is that in these Taylor approximants, for the time being, we can 
only incorporate next-to-leading order spin-orbit effects and all other 
spin effects are mainly at the `Newtonian order' (Newtonian order in the sense
one refers to radiation reaction appearing for the first time at 2.5PN (absolute) order  
as Newtonian radiation reaction).
Therefore, it is quite conceivable that smooth convergence to exact GW phase,
reported in Ref.~\cite{Gopakumar:2007vh} for the TaylorEt approximant for non-spinning 
compact binaries, would not be present in its spinning counterpart.

\section{NR-PN Comparison during inspiral}
\label{sec:pncomparison}

In this section we compare the phase and amplitude of the numerical waveforms
with that predicted by post-Newtonian approximants. We have
a lot of freedom in how to make such a comparison. Here
we follow the procedure we used
in \cite{Hannam:2007ik}, which is also used in \cite{Boyle:2007ft}: we line
up the phase and frequency of the PN and NR waveforms when the
gravitational-wave frequency $M \omega$ is equal to 0.1. This occurs several
orbits before merger (the number of orbits depends on the value of the
black-hole spin, as suggested by Table~\ref{tab:numresults}), at which time we
expect post-Newtonian results to still be 
valid. We then compare the phase and amplitude at earlier times. When
comparing results between different cases, we need to decide whether to
compare over a period of time, a frequency range, or a number of GW cycles. 
The results do not
change qualitatively with different choices; we simply must make some choice;
and the choice we make is to compare over a given number of GW cycles, which
it seems to us will be most useful when later constructing hybrid
waveforms. Our final procedure, then, is to compare the ten cycles before the
GW frequency reaches $M\omega = 0.1$. (This corresponds
roughly to a black-hole orbital frequency of $M\omega_b = 0.05$.)

\subsection{Phase}
\label{sec:phase}

\begin{figure}[t]
\centering
\includegraphics[height=4cm]{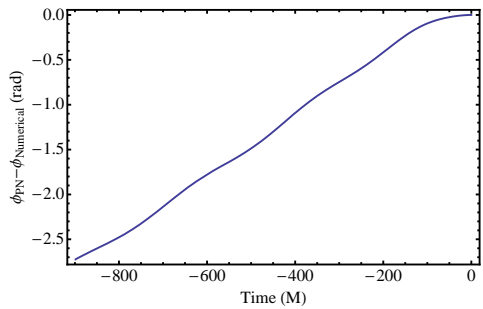}
\caption{An example of a phase comparison between NR and PN results. The case
  shown here is $S_i/M_i^2 = 0.5$, and the PN results were obtained with the
  TaylorT1 approximant at 2.5PN order. The accumulated phase error over the
  ten GW cycles before $M\omega = 0.1$ is $\Delta \phi = -2.69$~radians.}
\label{fig:PhaseComparisonExample}
\end{figure}

As an example of how our comparison procedure works,
Figure~\ref{fig:PhaseComparisonExample} shows the accumulated phase 
disagreement between the NR results for black holes with spins $S_i/M_i^2 =
0.5$, and the PN result using the Taylor T1 approximant at 2.5PN order. The
phase disagreement is zero by construction when $M\omega = 0.1$, and we choose
$t = 0$ at this point. As we progress backwards in time, the phase
disagreement at first grows quadratically (since $\dot{\phi}_{PN} =
\dot{\phi}_{NR}$ at $t = 0$, $(\dot{\Delta \phi}) = 0$, and so
there cannot be any linear growth in $\Delta \phi$ at ``early'' times). A
linear dependence soon develops, however, and the phase disagreement grows
roughly linearly for the remainder of the comparison, back to $t \approx
-900M$, when $\phi_{NR}$ has decreased by $20\pi$, or ten GW cycles. Note also
that there are wiggles in the plot; these are due to the small eccentricity in
the binary that was simulated numerically, and we expect that they introduce
an overall uncertainty in the numerical phase of about 0.2 radians with
respect to a binary with zero eccentricity (see Figure~15 in
\cite{Hannam:2007ik}).

\begin{figure}[t]
\centering
\includegraphics[height=4cm]{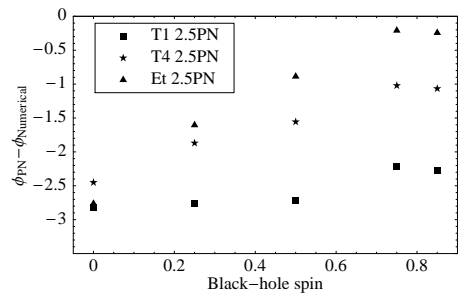}
\caption{The accumulated phase disagreement between NR and PN results over the
ten cycles before $M\omega = 0.1$. The five different spin values are
$S_i/M_i^2 = 0,0.25,0.5,0.75,0.85$, and we compare with the three approximants
TaylorT1, TaylorT4 and TaylorEt. All approximants are calculated at 2.5PN
order.} 
\label{fig:PhaseComparison25}
\end{figure}

\begin{figure}[t]
\centering
\includegraphics[height=4cm]{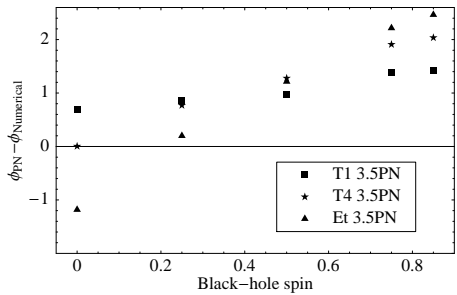}
\caption{The same comparison as in Figure~\ref{fig:PhaseComparison25}, but
  this time the PN approximants are evaluated at 3.5PN order in those terms
  where this is possible (see text).}
\label{fig:PhaseComparison35}
\end{figure}

In the example shown in Figure~\ref{fig:PhaseComparisonExample} the
accumulated phase error over the ten cycles before $M\omega = 0.1$ is $-2.69$
radians. This is the most important piece of information we obtain from the
plot, and rather than produce many very similar plots for different values of
black-hole spin compared with different PN approximants at different PN
orders, we summarize our results in Figures~\ref{fig:PhaseComparison25} and
\ref{fig:PhaseComparison35}.  Here
we see the accumulated phase agreement over ten cycles for five different
values of spin, ($S_i/M_i^2 = \{0,0.25,0.5,0.75,0.85\}$), compared with three PN
approximants, all calculated at 2.5PN and 3.5PN accuracy. The three approximants are
the standard Taylor T1 approximant, the Taylor T4 approximant introduced in
\cite{Boyle:2007ft}, and the Et approximant introduced in
\cite{Gopakumar07} and compared 
in detail with the nonspinning case in \cite{Gopakumar:2007vh}. The small eccentricity
tends to increase $\phi_{NR}$, so the points in Figures~\ref{fig:PhaseComparison25}
and \ref{fig:PhaseComparison35} are systematically too low by at most 0.2 radians
for the spinning cases,
and we should also recall the numerical uncertainty in $\phi_{NR}$ of 0.25 radians.

The main points
we wish to emphasize in Figure~\ref{fig:PhaseComparison25} are that the accumulated 
phase disagreement between NR
and 2.5PN TaylorT1 results is roughly constant for all values of black-hole
spin. This suggests that when producing hybrid waveforms for these cases, the
same number of numerical cyles are needed for all spin values to produce
hybrid waveforms of the same phase accuracy. On the other hand, the T4 and Et
approximants have the advantage that the phase disagreement is less than it is
for T1, and decreases for higher spins, with the Et approximant performing
best in the high-spin cases. 

Figure~\ref{fig:PhaseComparison35} shows a similar plot, but this time the
approximants are calculated up to 3.5PN accuracy. As pointed out in
Section~\ref{sec:PNmethods}, the 3.5PN results are not 3.5PN-accurate in all
terms, and these results will change when all 3.5PN terms are known and included.

We note once again that the phase disagreement between the NR and TaylorT1 phases
is roughly constant for all values of spins. We also see that, as
already seen in \cite{Boyle:2007ft}, the TaylorT4 approximant agrees extremely
well with the NR phase in the nonspinning case. However, this does not hold
for the spinning cases, and for larger spins 3.5PN TaylorT4 performs worse than
TaylorT1. The new TaylorEt approximant agrees extremely well in one case (when
$S_i/M_i^2 = 0.25$), but gives a large disagreement for high spins. 

Our general conclusion from all of these comparisons is that, for the very
small section of parameter space we have considered (equal-mass binaries with
non-precessing spins in the ``orbital hang-up'' configuration), the phase
agreement is roughly as good as in the nonspinning case; it certainly has not got 
dramatically worse, as one might have suspected. These numerical
simulations can therefore be used to produce hybrid waveforms of comparable phase
accuracy to those that can be produced from our previous nonspinning data
\cite{Hannam:2007ik,Ajith:2007qp,Ajith:2007kx,Ajith:2007xh}.

\subsection{Amplitude}

We now compare the amplitude of the $(l=2,m=2)$ mode of the NR $r\Psi_4$ with
that calculated from the restricted PN approximation, which is the same as the
amplitude from the quadrupole formula. The restricted PN amplitude of the GW
strain $h$ is given by $A_{\rm restricted} = (M\omega/2)^{2/3} / R$, where $R$
is the distance from the source, and we have \begin{equation}
h = A_{\rm restricted} e^{i \phi_{PN}(t)}.
\end{equation} The phase $\phi_{PN} (t)$ depends on the PN approximant used in
the previous section. However, if we differentiate $h$ twice with respect to
time to calculate $r\Psi_{4,22}$, and consider the amplitude as a function of GW
frequency $M\omega$, then we find that the choice of approximant has a
negligible effect on the resulting function. In other words, the GW strain
amplitude $A_{\rm restricted}$ depends only on the GW frequency, and so does
the restricted PN amplitude of $r\Psi_{4,22}$, to a good approximation. This
point was also made in \cite{Boyle:2007ft}. This fact allows us to make a
comparison between the NR and restricted PN amplitudes that is independent of
the PN approximant that we used to calculate the phase --- the results are
independent of whether we use Taylor T1, T4, or Et. 

In the nonspinning case we found that the NR and restricted PN amplitudes
disagree by about $(6 \pm 2)$\% \cite{Hannam:2007ik}, while \cite{Boyle:2007ft}
found 5\% disagreement using numerical waveforms with higher accuracy. Both
sets of numerical results found that the disagreement was roughly constant
over the last ten cycles before $M\omega = 0.1$. 

\begin{figure}[t]
\centering
\includegraphics[height=4cm]{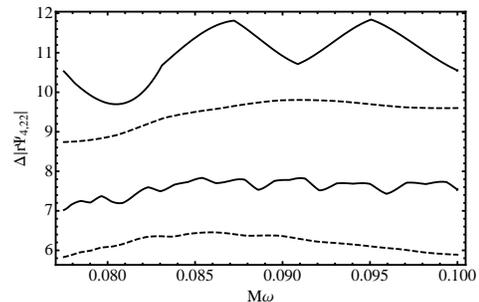}
\caption{The disagreement between restricted NR and PN amplitudes of $r\Psi_{4,22}$ 
as a function of GW frequency  $M\omega$. The lines
  below correspond, from bottom to top, to the cases $S_i/M_i^2 =
  0,0.25,0.5,0.75$. (Every second line is dashed, to make them easier to distinguish.)
In the nonspinning case the disagreement
is roughly 6\% over this frequency range. The disagreement increases as spin
is added, and is about 11\% for $S_i/M_i^2 = 0.75$. The large oscillations in this 
last case are due to the relatively high eccentricity of that system, $e \sim 0.006$.}
\label{fig:AmplitudeComparison}
\end{figure}

The results for the spinning binaries we have considered are shown in
Figure~\ref{fig:AmplitudeComparison}. We see that for spinning binaries the
amplitude disagreement increases. The plot shows results for spins $S_i/M_i^2
= 0,0.25,0.5,0.75$. For the highest spin shown here, the amplitude
disagreement between NR and restricted PN is about 11\%. In the case of
$S_i/M_i^2 = 0.85$, fluctuations due to eccentricity make it difficult to
clearly measure the amplitude disagreement, and that case is not shown in the
plot; however, we estimate the disagreement at around 12\%, which is twice as
high as in the nonspinning case. 

Higher-order PN amplitude corrections were able to improve the agreement in
the nonspinning case to around 2\% using 2.5PN amplitude corrections
\cite{Hannam:2007ik,Boyle:2007ft}, and even better agreement was found when
using 3PN amplitude corrections \cite{Boyle:2007ft}. It is clear from our
current results that higher-oder PN amplitude corrections are even more
crucial when spins are included. Higher-order corrections up to 2.5PN already
exist \cite{Kidder1995,Blanchet:2006gy,Faye:2006gx}, but 3PN amplitudes are
not yet known for spinning binaries. These corrections may substantially
improve the agreement with NR amplitudes, but we will postpone that study to
future work. 

\section{Inspiral and merger}
\label{sec:numcomparison}

We may ask how much the waveforms from spinning and nonspinning binaries
differ. In data analysis applications the overlap integrals
between different waveforms determine how well a
GW search will distinguish between them. If the phase of two waveforms can
be lined up such that the waveforms are identical, then the overlap between
them is one. If the waveforms differ slightly, the value of the overlap
integral falls below one. In real-world data-analysis applications, the detector
noise is included in the overlap integral, which leads to results that depend
on the total mass of the binary.
For the purposes of the purely qualitative
illustration in this section, we do not use the detector noise, and we
calculate the overlap integrals in the time domain. 
Given some part of the waveform, we may calculate an overlap integral of the form,
 \begin{equation}
O(\delta t,\delta \phi) = \frac{1}{N_{NS} N_S} \int_{t_1}^{t_2} h_{NS}(t) 
h^{\star}_S(t + \delta t,\delta \phi) dt, \label{eqn:overlap}
\end{equation}
where ``NS'' indicates the waveform from a
nonspinning binary, and ``S'' indicates one of the spinning-binary
waveforms. The spinning waveform is time-shifted by some $\delta t$, and
phase-shifted by some $\delta \phi$. The normalization factors are calculated
by performing the integrals  \begin{eqnarray}
N^2_{NS} & = & \int_{t_1}^{t_2} | h_{NS}(t) |^2 dt \,,\\
N^2_{S} & = & \int_{t_1}^{t_2} | h_{S}(t) |^2 dt.
\end{eqnarray}
For some choice of $t_1$ and $t_2$, the time and phase shifts
$\delta t$ and $\delta \phi$ can be optimized to find the largest possible
overlap. 

Let us compare the nonspinning-binary waveform with that from a binary with
$S_i/M_i^2 = 0.75$. We first choose a time interval that includes only
inspiral, $t_1 = 1200M$ and 
$t_2 = 1700M$ with respect to the nonspinning waveform; this corresponds to
roughly five GW cycles in the frequency range $M\omega \approx 0.06$ to
$M\omega \approx 0.09$. The
maximum overlap that can be achieved when comparing this
small number of cycles with the S75 simulation is
just under 0.99, meaning that 
the overlap is extremely good. This illustrates that during the inspiral, it is
difficult to distinguish between the spinning and nonspinning waveforms. Put
another way, if one of these waveforms were detected, a large number of cycles
would be required to
estimate the black holes' spins. In
practice, indeed many hundreds or thousands of cycles may be detected during the
inspiral, and the accumulated phase difference between spinning and
nonspinning binaries may be more obvious. However, at present we must use PN
approximants to model those hundreds or thousands of cycles, and
it is not entirely clear how the variation
in number of cycles when spin is added compares with the deviation of
the PN approximants at different PN orders from the true solution.

In addition, if a noticeable phase difference between spinning and nonspinning 
waveforms does accumulate over many hundreds of
cycles, this might be partially compensated by rescaling the total mass of one of the 
binaries;
a spinning binary may thus be detected as a nonspinning binary 
with the wrong mass. 

We now examine a time interval that includes the merger, $t_1 = 1870M$ and
$t_2 = 2020M$. The maximum overlap that we can now achieve is less than 0.90:
the waveforms differ far more in the merger phase. This is a far greater
difference than we observe during the inspiral, and a far greater difference
than we would expect to see even if many more inspiral cycles were available. 

Furthermore, if we were to adjust the total mass of one of the binaries, such that 
the overlap of the inspiral waveforms was increased to almost one, this would
{\it not} significantly improve the overlap of the merger waveforms. Therefore,
detecting both the inspiral {\it and} merger may allow a significantly better
estimate of the black holes' parameters. 

\begin{figure}[t]
\centering
\includegraphics[height=4cm]{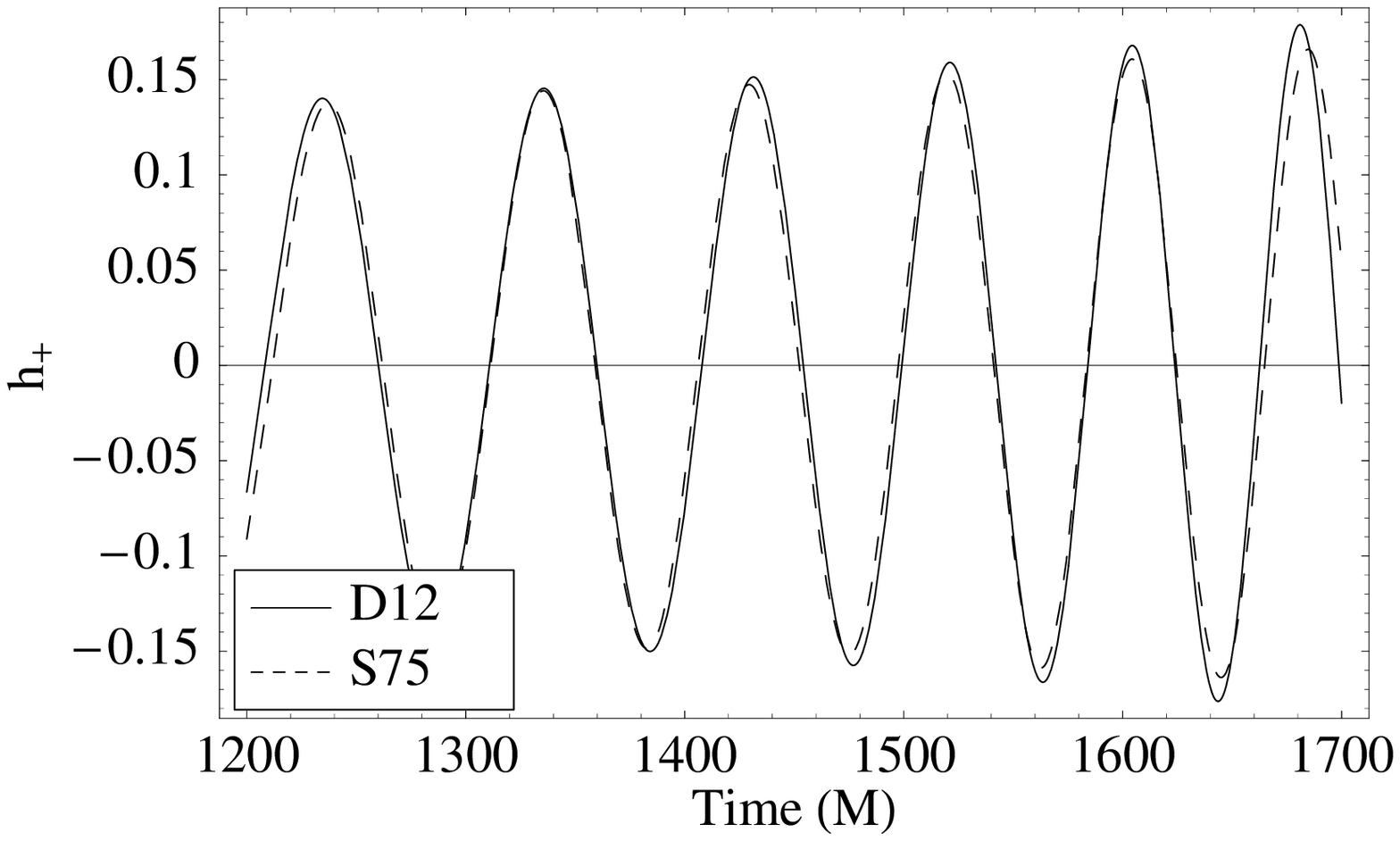}
\includegraphics[height=4cm]{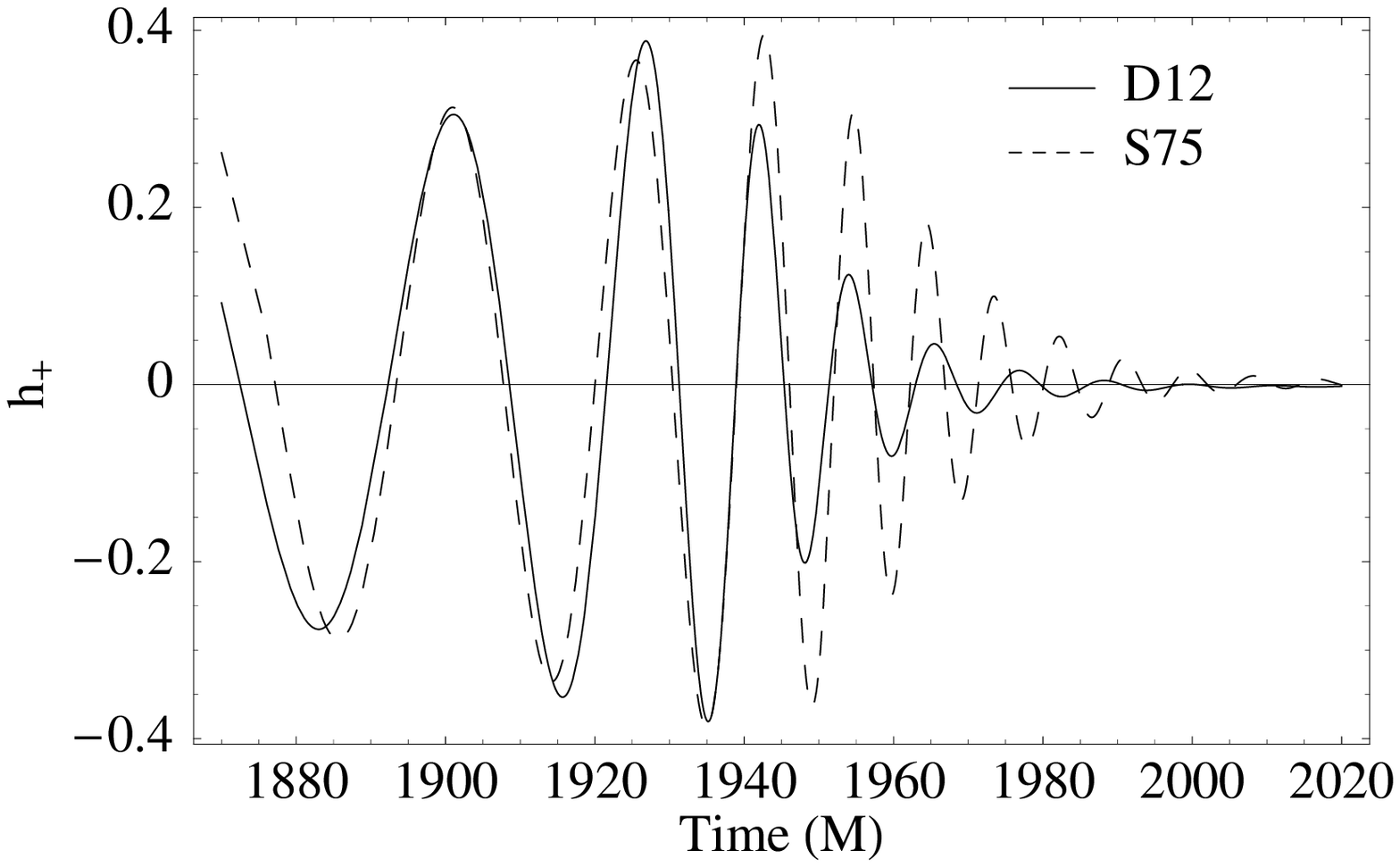}
\caption{Spinning and nonspinning waveforms in the time intervals
  $t=[1200,1700]$ and $t=[1870,2020]$, with the time and phase shifted so that
they give the maximum value of the overlap integral (\ref{eqn:overlap}). We
can clearly see by eye that the overlap for the earlier time interval is much
larger than that for the later interval.}
\label{fig:overlap}
\end{figure}

In Figure~\ref{fig:overlap} we illustrate this point by showing the waveforms
for the nonspinning and 
$S_i/M_i^2 = 0.75$ simulations, over the two time intervals we just discussed,
and lined up such that the overlap integral (\ref{eqn:overlap}) is a
maximum. It is immediately clear from the two plots that the overlap will be
larger for the earlier time interval. 

We conclude from this simple analysis that, at least for binaries in an
orbital hangup configuration, an accurate estimate of the black holes' spins will
be far easier if the merger waveform is detected, rather than the inspiral. And,
perhaps not surprisingly, the most accurate parameter estimation would be
possible if both the inspiral and merger were detected.  A similar conclusion can be
made when looking at higher modes, although all modes up to $l=4$
have an amplitude no more than 10\% that of the $(l=2,m=2)$ mode in the cases
we consider.

\section{Discussion}

We have extended our previous comparison of NR and PN inspiral waveforms
\cite{Hannam:2007ik} to spinning binaries. In particular, we considered
equal-mass binaries whose black holes have equal spins oriented parallel to
the orbital angular momentum. In these cases the spins do not precess and
there is no net radiated linear momentum, but there is an ``orbital hangup''
effect that delays the merger as the spin is increased
\cite{Campanelli:2006uy}. 

We compared the accumulated phase difference between NR and PN
approximants over the ten cycles before the GW frequency reaches $M\omega =
0.1$. We used the PN approximants TaylorT1, T4 and Et, at orders 2.5PN and
3.5PN. We conclude that, as in the nonspinning case, the growth of phase error
is fairly small; for the number of cycles we considered, it is less than 2.5
radians for all 3.5PN approximants for spins up to $S_i/M_i^2 = 0.85$, and
less than 1.5 radians for the TaylorT1 approximant. The 3.5PN
TaylorT4 approximant, which performed extremely well in the equal-mass case
\cite{Boyle:2007ft} is seen to perform much worse in the spinning cases. However,
we stress that not all terms in the PN approximants are known to 3.5PN
order. At 2.5PN order (the highest at which all terms are known), the Et
approximant gives the best phase agreement. These results are summarized in
Figures~\ref{fig:PhaseComparison25} and \ref{fig:PhaseComparison35}. 

We also compare the amplitude of the $(l=2,m=2)$ mode of the NR $r\Psi_4$ with
that calculated by the restricted PN approximation. In the equal-mass case the
PN amplitude was larger than its NR counterpart by about 6\%. 
In a binary whose black holes have spins $S_i/M_i^2 = 0.85$, the amplitude disagreement
grows to around 12\%. 

Finally, we gave an illustration of how the inspiral and merger
waveforms differ between the spinning and nonspinning cases. Our simple
analysis suggests that the spin is difficult to distinguish during inspiral,
but has an extremely clear effect on the merger waveform. As such, we expect
that estimating the spins of a binary's constituents should be much easier in
cases where the merger is also detected.

In the future we intend to extend this analysis to a larger sample of
parameter space (in particular, more general spin configurations), and to explore
the data-analysis implications of our results.

\acknowledgments

AG is grateful to Gerhard Sch\"afer for
fruitful discussions and Manuel Tessmer for 
crosschecking PN formulae, and SH and MH for discussions with Alicia Sintes.
This work was supported in part by
DFG grant SFB/Transregio~7 ``Gravitational Wave Astronomy''
and the DLR (Deutsches Zentrum f\"ur Luft- und Raumfahrt) through ``LISA 
Germany''. We thank the DEISA Consortium (co-funded by the EU, FP6
project 508830), for support within the DEISA Extreme Computing Initiative
(www.deisa.org). Computations were performed at LRZ Munich and the Doppler and
Kepler clusters at the Institute of Theoretical Physics of the University of
Jena.

\bibliography{refs}

\end{document}